# Possible Class of Nearby Gamma-Ray Burst / Gravitational Wave Sources

Jay P. Norris

*NASA/Goddard Space Flight Center*
*Laboratory for High Energy Astrophysics*

**Abstract.** A possible subclass of gamma-ray bursts – those with few, wide pulses, spectral lags of order one to several seconds, and soft spectra – has been identified. Their Log[N]-Log[Fp] distribution approximates a −3/2 power-law, suggesting homogeneity and relatively nearby sources. These mostly dim bursts account for ~ 50% of the BATSE sample of long bursts near that instrument's trigger threshold, suggesting that this subluminous class constitutes a more common variety than the more familiar burst sources which lie at truly cosmological distances. Theoretical scenarios predicted such a class, motivated by their exemplar GRB 980425 (SN 1998bw) lying at a distance of ~ 38 Mpc. The observations are explained by invoking off-axis viewing of the GRB jet and/or bulk Lorentz factors of order a few. Long-lag bursts show a tendency to concentrate near the Supergalactic Plane with a quadrupole moment of −0.10 ± 0.04, similar to that for SNe type Ib/c within the same volume. The rate of the observed subluminous bursts is of order ¼ that of SNe Ib/c. Evidence for a sequential relationship between SNe Ib/c and GRBs is critiqued for two cases, as simultaneity of the SN and GRB events may be important for detection of the expected gravitational wave signal; at most, SN to GRB delays appear to be a few days. SN asymmetries and ultrarelativistic GRB jets suggest the possibility of rapid rotation in the pre-collapse objects, a primary condition required for highly non-axisymmetric SN collapse to produce strong gravitational waves.

## INTRODUCTION

The vast majority of gamma-ray burst (GRB) sources for which associated redshifts have been obtained – by examination of either the spectrum of the optical afterglow or that of the host galaxy – lie at cosmological distances, as shown in Figure 1. However, the two bursts with the lowest determined redshifts (z = 0.0085, 0.1685) may bracket a subpopulation of subluminous GRBs, which may actually dominate the numbers in the parent population. As described in the next section, in the view of most of the GRB/SN community these two bursts have been shown to be definitively associated with high-mass progenitor, core-collapse supernovae (SNe). This solid GRB-SN connection and the possibility that massive SNe might collapse in highly non-axisymmetric modes forms the basis for discussion of detectable gravitational waves (GW) from these events. The sparse evidence that the GRB and associated SN events are (nearly) simultaneous is also discussed. In the succeeding section, attributes of the long-lag GRB subpopulation are described, supporting the case that the detected GRBs in this ultra-low luminosity group lie within distances of ~ 100

Mpc, comparable to the distance scale for virtually all detected SNe type Ib/c. The final section discusses the central issue – the maximum strain from this putative class of GRBs/SNe sited at the presumed distances, given the assumption of a very optimistic non-axisymmetric deformation of order unity during SN collapse.

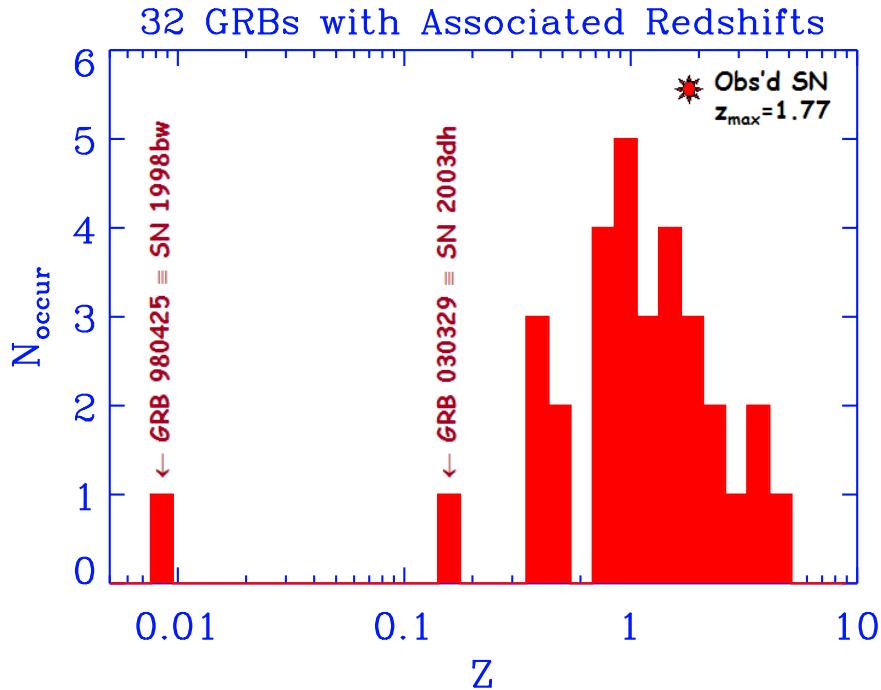

**FIGURE 1.** Distribution of redshifts associated with GRB afterglows and their host galaxies. For the two GRBs with lowest redshifts, and lowest luminosities, evidence of SNe light curves and spectra were found with the GRB afterglows. GRB 980425 is the canonical example of a subgroup of GRBs with long spectral lag, believed to have mild to subrelativistic jets; GRB 030329 may represent the transition to ultrarelativistic jets. Most GRBs with determined redshifts lie at cosmological distances, with median redshift $z \sim 1$. For reference, the redshift of the most distant SN so far detected is depicted.

## SUPERNOVA — GAMMA-RAY BURST RELATIONSHIP

Historically, SNe were first classified into two groups according to the presence (type II) or absence (type I) of discernible hydrogen emission lines [1; see their Figure 2]. Physically, however, the two mechanisms leading to explosion are believed to be accretion-induced collapse of a white dwarf (type 1a) or core-collapse following cessation of nuclear burning (all type II, type Ib and Ic). Essentially, for the core-collapse events, increasing progenitor mass determines the observed SN spectrum, which corresponds to increasing envelope mass loss with SN species: II (H lines), Ib (no Si lines, but He present) or Ic (no He lines).

The pair of events which initiated the GRB/SN connection were the ultra-low luminosity ($\sim 10^{47}$ erg s$^{-1}$) GRB 980425, and the type Ic SN 1998bw. The SN lay within the BeppoSAX X-ray error circle of 1 arc minute radius determined for the GRB [2], and the SN onset was modeled to lie within +0.7/–2.0 days of the burst [3]. The expectation for such a chance occurrence was estimated to be ~ $10^{-4}$ [4]. The SN observations indicated unusually high $^{56}$Ni production as well as high radio emission, with the progenitor modeled as a 13.8 $M_\odot$ C+O star [3,4]. SN 1998bw lies at a distance of ~ 38 Mpc (z = 0.0085), similar to the median distance for detected type Ib/c SNe, as shown in Figure 2. The sky distribution for type Ib/c SNe is plotted in Figure 3 in the Supergalactic (SG) coordinate system [5], whose plane corresponds to the somewhat flattened distribution of galaxy clusters within ~ 100 Mpc. The SNe distribution is also oblate, with quadrupole moment Q = –0.110 ± 0.037, reflecting their sites within nearby galaxies.

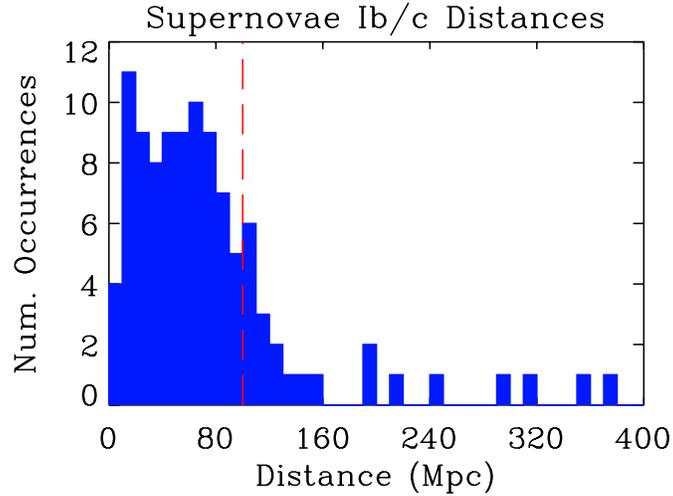

**FIGURE 2.** Distribution of distances for all detected type Ib and Ic SNe, 1954–2003.4. Dashed line at 100 Mpc demarcates rough limit within which nearby galaxy superclusters form a somewhat flattened distribution, the reference for the Supergalactic coordinate system. (Assumes $H_0 = 72$ km s$^{-1}$ Mpc$^{-1}$. Three SNe at distances ~ 1.5 Mpc, discovered in type Ia searches, are not included). SNe data are from the "List of Supernovae," http://cfa-www.harvard.edu/iau/lists/Supernovae.html (positions) and the NASA/IPAC Extragalactic Database, http://nedwww.ipac.caltech.edu/forms/byname.html (distances).

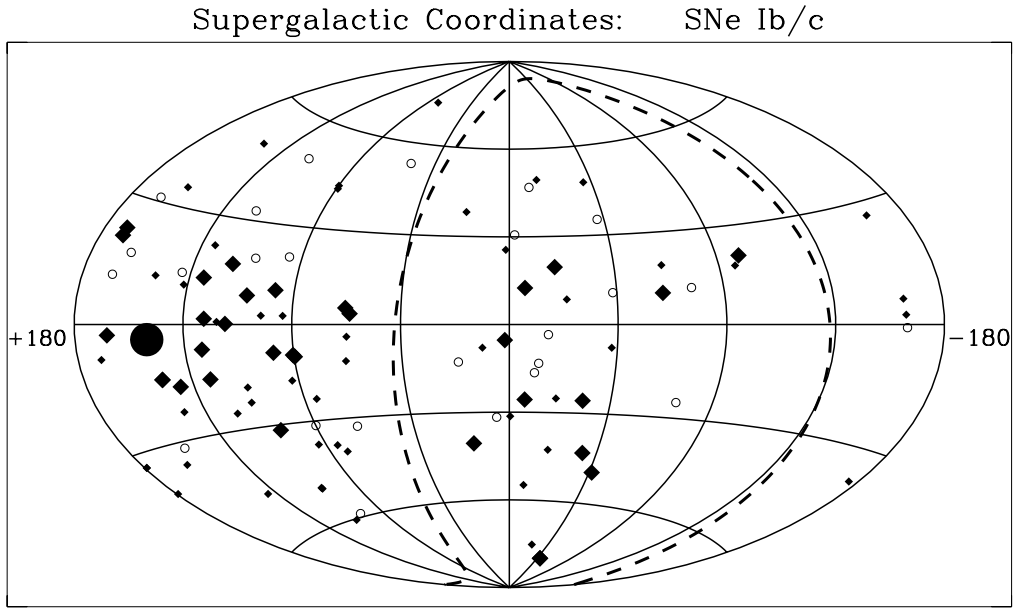

**FIGURE 3.** The sky distribution for all detected SNe Ib, Ic in the Supergalactic coordinate system. Symbols indicate distances: large diamonds (d < 40 Mpc), small diamonds (40 Mpc < d < 100 Mpc), and open circles (d > 100 Mpc). Large filled circle depicts direction of center of Virgo supercluster. For SNe with d < 40 Mpc (100 Mpc), the quadrupole moment with respect to the Supergalactic Plane, $Q = -0.170 \pm 0.053$ ($-0.110 \pm 0.037$). Dashed great circle represents the Galactic Plane.

A second convincing event pair, GRB 030329 / SN 2003dh, was determined from the GRB afterglow spectrum to lie at a distance of ~ 800 Mpc (z = 0.1685), about 20 times further than the 1998 GRB/SN pair. The afterglow revealed a bump on the timescale of ~ 10–60 days after GRB onset, consistent with the radioactivity peak of a SN. This evidence in itself is not convincing proof of the presence of a SN since other viable explanations for (often ill-defined) bumps and wiggles in GRB afterglow time profiles have been advanced for GRBs at higher redshifts. Also, type Ic events show at least a factor of ~ 3 variation in the full-width half maximum of the radioactivity peak, and this variation does not necessarily scale monotonically with the modeled $^{56}$Ni production [3; see their Figure 1]. However, the spectrum of SN 2003dh does look convincingly similar to that of SN 1998bw [6]. Interestingly, the high flux at Earth ranks GRB 030329 amongst the top ~ 1% of bursts in brightness, but given the relatively close distance, its rank in peak luminosity is near the low end for bursts with determined redshifts, several × $10^{51}$ erg s$^{-1}$. The luminosities quoted here for GRBs 980425 and 030329 are for a *would-be isotropic* geometry, i.e., a spherically beamed emission pattern, whereas the jetted outflows with opening angles of ~ 2–20° that are inferred for highly luminous bursts, $10^{51-54}$ erg s$^{-1}$, render the 4π-integrated luminosities much lower and more uniform [7,8,9,10]. The two under-luminous GRBs described here presumably manifest relatively wider jet opening angles in their relativistic flows [11,12]. Small (large) opening angles may be related to high (low) angular rotation rate [13], and to the potential for generating instabilities during core collapse and thence non-axisymmetric bar configurations that might produce detectable GW radiation [14,15]. Thus, the lower potential for bar instabilities to develop in SN collapses associated with such under-luminous bursts may offset to some (unknown) degree the GW-detection advantage afforded by their closer distances. The next section will discuss attributes of the subpopulation to which GRBs 980425 and 030329 appear to belong, arguing for distances of < 100 Mpc for the majority of these bursters. Before proceeding, we examine the available evidence for a two-event sequence, SN followed by GRB, and briefly summarize the evidence for asymmetric (polar) SN explosions.

The issue of whether the SN happens first (core-collapse) followed by the GRB (residual accretion) – or if the two events are simultaneous – relates to strength of the two (or one) GW signals and to detectability: The accuracy of the time marker provided by the GRB is definitive, whereas the exact time of SN core collapse is never observed and is usually uncertain by days. Thus, many more GW search trials would need to be performed surrounding the predicted SN time. Some scenarios involve such a two-stage process with the timescale between collapse and explosion governed, for instance, by accretion and other torques, and black hole spin rate [16, 17].

Evidence from two bursts suggests that $T_{0\_GRB} - T_{0\_SN}$ was, at most, of order a few days. The light curve for SN 1998bw in two colors [4; see their Figure 1] shows an upturn (defined by two points in time) prior to the broad radioactivity peak, at less than one day past GRB 980425 onset. As mentioned above, the total SN light curve was modeled to yield a SN onset within +0.7/–2.0 days of the burst [3], consistent with SN-GRB simultaneity. However, this apparent minimum could be interpreted as following the rarely observed "ultraviolet (UV) breakout" maximum. SN 1993J (a type II) shows an interval of ~ 8 days from onset to minimum between the two maxima [18; see their Figure 2]. In this fortuitous case, even the steep rise through ~ 8 magnitudes to the UV breakout maximum was observed; hence the time of SN onset is relatively accurately constrained. In view of the SN 1993J behavior, the (unobserved) UV maximum in SN 1998bw, if indeed there was one, could conceivably have occurred several days earlier than the GRB. The second case involves the interpretation of the X-ray spectrum of the afterglow of GRB 011211 in terms of early emission lines expected in SN ejecta, where the lines were measured to be blue-shifted with respect to the GRB redshift by v/c ≈ 0.09. Reasoning from particular assumptions on jet opening angle and interstellar density, the conclusion was that the GRB ejecta, traveling near the speed of light, collided with the SN shell at R ~ $10^{15}$ cm, ejected ~ 4 days previously [19]. Other interpretations may be possible, and the inference of a SN-GRB delay rests on the X-ray counting statistics (ranging from modest to poor), which define the putative emission lines.

The essential ingredient for LIGO-detectable GW radiation from SNe-GRBs is the formation of a strong non-axisymmetric "bar" instability during core collapse. The primary prerequisite for non-axisymmetric deformation is rapid core rotation [15]. Several lines of evidence indicate that some SNe explode asymmetrically preferentially along the poles. Early indirect evidence, high pulsar space velocities, implied net momentum presumably resulting from instabilities that develop during rotational core collapse. More recently, images of SN 1987A have been interpreted in terms of polar and equatorial asymmetries that are resolved as the remnant expands [20]. Elemental asymmetries in older remnants such as Cassiopeia A also argue for asymmetric nuclear synthesis and polar ejection, and SN modeling supports this picture of strong polar asymmetry [20]. Closest to the event itself are polarization observations which show clear trends. For type Ia SNe, polarizations of ~ 0.3% are typical; whereas for the more energetic type II, polarizations of 1–2% are observed to increase with time as the ejecta expand, revealing the more asymmetric core material. The most energetic core-collapse SNe – type Ib/c with less overlying shell material at onset – exhibit polarizations of 3–7% [21].

GRBs may reveal a more complete picture of the dynamics of the SN core collapse event (if in fact, $T_{0\_GRB} \approx T_{0\_SN}$). The first measurement of gamma-ray polarization was made recently with *RHESSI* for GRB 030329. The event-averaged hard X-ray emission had an astonishing 80±10% polarization (much higher than any other astrophysical source so far observed), near the theoretical maximum, strongly suggesting a highly ordered magnetic field that arises near the source [22]. Theoretical interpretations of this high degree of polarization argue that the explosive outflow is likely to be magnetically rather than hydrodynamically dominated, with the observed emission arising from a large-scale, causally connected region, and the primal energy coming from a relativistically rotating stellar-mass progenitor [13,22]. The picture of a rotation-driven, magnetically dominated jet is reinforced by the interpretation of breaks in the power-law decays of GRB afterglows as transitions to subrelativistic flow, where the jets are originally beamed into solid angles of ~ $4\pi/(\text{few}\times 100)$ [7,8,9,10]. Thus the primary condition required for non-axisymmetric (bar) instabilities near unity to develop – high T/W: kinetic/potential energy [14] – i.e., rapidly rotating pre-collapse stellar cores, might obtain in SN-GRB events.

## A NEARBY POPULATION OF SUBLUMINOUS GRBS

In the last four years efforts by several groups have discovered luminosity measures for GRBs with associated redshifts. In principle, these relationships, based solely on gamma-ray characteristics, may then be used to estimate coarsely the luminosities and distances for larger samples without benefit of redshifts. The anti-correlation between average spectral lag and luminosity [24] will be discussed here in the context of the subset of BATSE bursts with very long lags [25]. In addition, temporal variability [26,27] and peak in the $\nu \cdot F(\nu)$ spectrum [28] have been shown to be robust measures directly correlated with burst luminosity. Since GRB jet models have at least two degrees of freedom (e.g., Lorentz factor and viewing angle), it is reasonable to expect that a combination of spectral and temporal measures might eventually produce more accurate measures of GRB luminosity.

The GRB subpopulation which may have the highest potential for GW detection has defining gamma-ray characteristics at the temporal and spectral extremes of long GRBs ($T_{90}$ duration > 2 s). Nearly all intermediate- to intense-flux GRBs have many, narrow pulses. However, bursts with a few, very long pulses and long spectral lag ($\gtrsim$ 0.3 s) between low and high energy bands (25–50 keV vs. 100–300 keV) comprise about half of the long-duration BATSE bursts near trigger threshold [25; see their Figure 3]. These "long-lag" bursts have soft spectra, but durations that are similar to the those of the main population [29]. Figure 4 illustrates the defining temporal differences between long and short lag bursts. "Subluminous" GRB 980425 is the canonical example of a long-lag burst ($\tau_{\text{lag}} \approx 2.8$ s), similar in appearance to the burst shown in the top panel of Figure 4, but with only one wide pulse.

In Figure 5 the BATSE localizations for relatively long-lag ($\tau_{\text{lag}} > 2$ s) bursts are plotted in the SG coordinate system, as in Figure 2 for the detected SNe Ib/c population. The tendency to concentrate towards the SG plane – defined by the superclusters within ~ 100 Mpc – is significant at about the 2.5-$\sigma$ level, with an oblate quadrupole moment Q = –0.097 ± 0.038. The flattened sky distribution for long-lag bursts and the ~ 38 Mpc distance for its canonical member, GRB 980425 / SN 1998bw, suggest that this subpopulation is closer by two orders of magnitude than the vast majority of GRBs with known redshifts (Figure 1). Unlike these short-lag bursts at cosmological distances, whose integral peak flux–frequency distribution (Log $N[>F_p]$ – Log $F_p$) turns over due to General Relativistic effects, the distribution for long-lag bursts follows a –3/2 power-law over a factor of ~ 40 in peak flux, suggesting a homogeneous space density unaffected by cosmology [25; see their Figure 4].

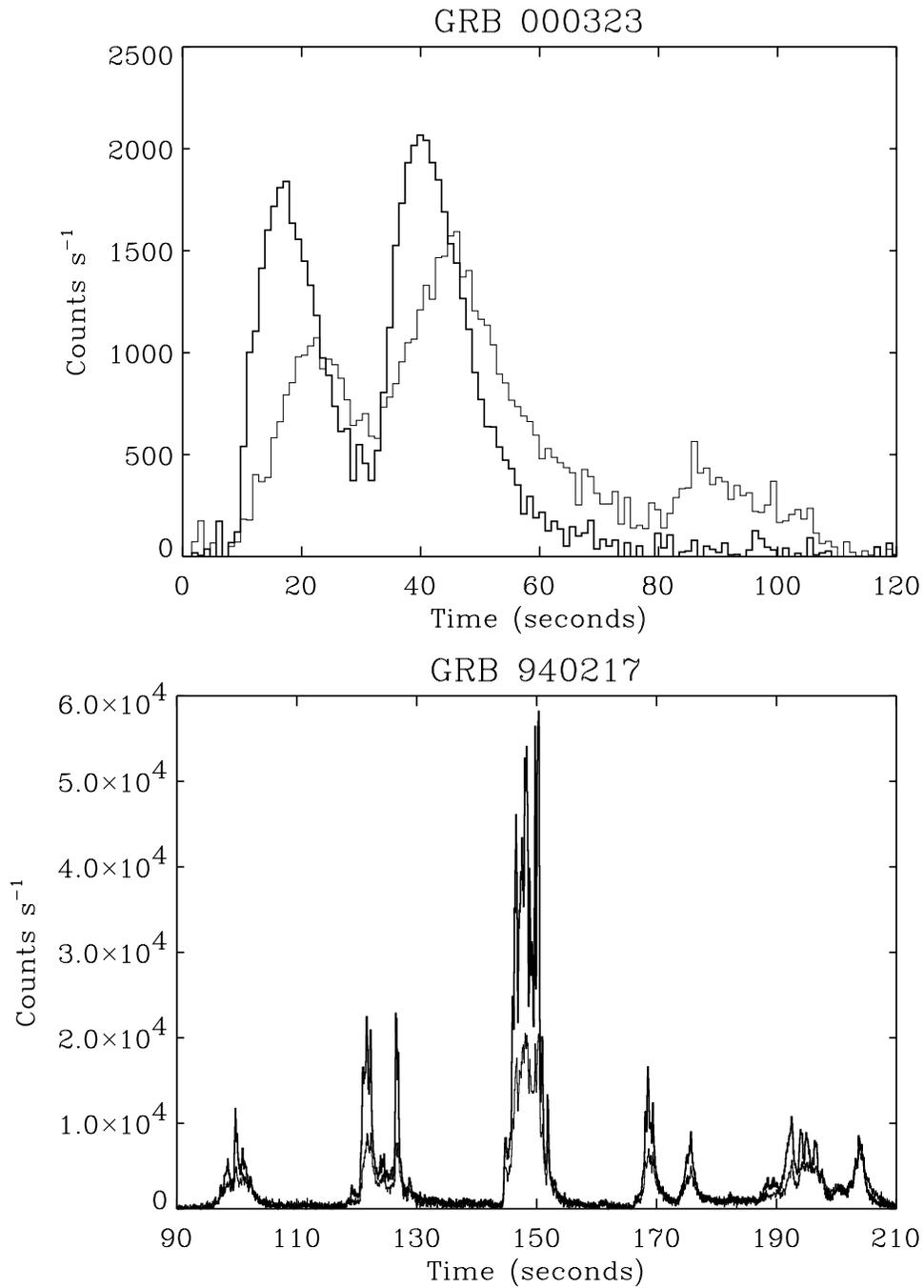

**FIGURE 4. Top:** Example of long-lag burst in two energy channels, 100–300 keV (thick line) and 25–50 keV (thin line); spectral lag ~ 4 s. Long-lag bursts characteristically have few, wide pulses. **Bottom:** Major portion of typical short-lag burst, on same timescale as top panel. Each of the several emission episodes consists of several narrow pulses; average spectral lag for this burst is ~ 20 ms.

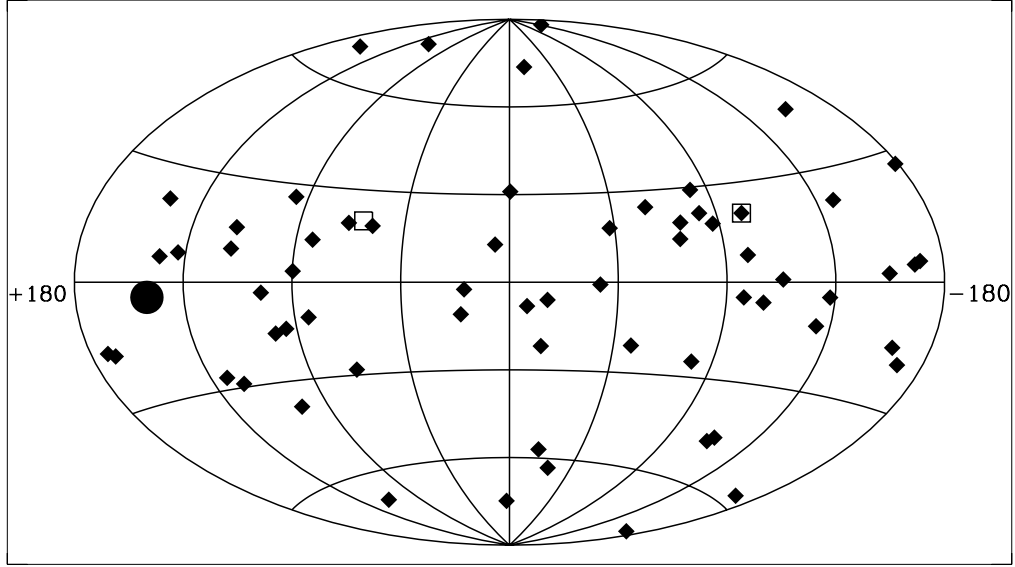

**FIGURE 5.** The sky distribution for long-lag gamma-ray bursts, with threshold $\tau_{lag} > 2$ s, in the Supergalactic coordinate system. Diamonds indicate the long-lag sample. Diamond inside square is GRB 980425. Empty square denotes GRB 971208 – longest, single-pulse BATSE burst; lag ~ 20 s. For this subset of bursts, $Q = -0.097 \pm 0.038$. Large filled circle indicates center of Virgo supercluster.

If this picture of a nearby subluminous GRB population is correct, then Universe-wide the GRB rate-density would be dominated by undetected long-lag GRBs: We could see the luminous, long-duration BATSE bursts to redshifts as high as $z \sim 30$ [30] at a rate of ~ 500 yr$^{-1}$. Whereas, ~ 7% of bursts have long lags – about 30 yr$^{-1}$ when corrected for BATSE exposure – but they are detected to distances of only ~ 100 Mpc, or ~ $10^{-5}$ × volume of Universe, neglecting galaxy density evolution effects. This implies ~ $3 \times 10^6$ long-lag bursts yr$^{-1}$ Universe$^{-1}$. These numbers are consistent with the SN Ib/c rate estimated at ~ 0.3 day$^{-1}$ within 100 Mpc [31] (only ~ 1/10 of nearby SNe are detected), since ~ ¼ of these highly energetic SNe would be required to generate the subluminous GRBs. These numbers are derived from an assumed two-branch lag-luminosity relationship [24; see their Figure 6], where the lower branch is defined only by GRB 980425 and the theoretical prediction of a transition [32] to subrelativistic flow for GRB luminosities $\lesssim 10^{51}$ ergs s$^{-1}$.

For GRB 030329 / SN 2003dh to support the two-branch relationship, this second closest GRB with a determined redshift (z = 0.1685) should lie slightly above the transition, with a spectral lag of ~ 0.1–0.2 s, measurable by HETE-2. A larger fraction of soft-spectrum, long-lag bursts is expected to be detected by Swift since it is more sensitive than BATSE, and its trigger energy threshold will most likely extend below that of BATSE (25 keV), to perhaps as low as 10 keV. Presumably, redshifts for long-lag bursts have not been determined in greater numbers by BeppoSAX and HETE-II since these instruments are slightly less sensitive than BATSE [33], tending to miss the dimmer bursts near BATSE threshold, and afterglows from subluminous bursts probably tend to be shorter lived and dimmer. Interestingly, theoretical predictions of a numerous subpopulation of subluminous GRBs (accommodating GRB 980425) directly preceded [11,12] their characterization observationally. The soft-spectrum and long-lag attributes are predicted to arise from mild-to-subrelativistic outflows, wide opening angles, and/or large viewing angles [32,34,35]. Variations on GRB jet opening angles and profiles are discussed in several recent articles [7,8,9,10].

## GRAVITATIONAL WAVE SIGNAL FROM NEARBY GRBS?

For the long-lag GRB sources to produce GW detectable by LIGO-II would require development of bar instabilities with T/W approaching unity [14,15]. While such deformations have not been demonstrated in core-collapse SN simulations, we can still speculate on detectable GW rates assuming these deformations are produced. Within 50 Mpc, the long-lag GRB rate would be ~ 4 yr$^{-1}$. At this distance, the strain in the 200–800 Hz band assuming a duration of 100 cycles would be h/√Hz > $1.3 \times 10^{-23}$, a factor of several above the anticipated sensitivity of LIGO-II [36]. At the distance of GRB 030329 (~ 800 Mpc), the maximum strain would be near the LIGO-II threshold. Of course, the distinct advantage that GRBs have over other potentially detectable GW sources is the electromagnetic signal which provides an accurate time marker, and thus the number of search trials is considerably reduced – if in fact SN and GRB are simultaneous or nearly so.

The main caveat in this argument is that the putative nearby GRB population is subluminous, which may be partially attributable to relatively low bulk Lorentz factors, in turn related to lower rotation rates in the pre-collapse object [15]. Thus, lower potential for bar instabilities to develop probably offsets to some (unknown) degree the GW-detection advantage afforded by the nearby sources.

## ACKNOWLEDGMENTS

I would like to thank David Band, Jerry Bonnell, Jordan Camp, John Cannizzo, Joan Centrella, and Timothy Young for many informative conversations concerning supernovae, gamma-ray bursts, and gravitational wave detection.

# REFERENCES

1. Cappellaro, E., & Turatto, M., "Supernova Types and Rates," in *The influence of binaries on stellar population studies*, Dordrecht: Kluwer Academic Publishers, 2001, xix, 582, p. 199
2. Pian, E., et al., *A&AS* **138**, 436 (1999)
3. Iwamoto, K., et al., *Nature* **395**, 672 (1998)
4. Galama, T. J., et al., *Nature* **395**, 670 (1998)
5. Hudson, M.J., *MNRAS* **265**, 43 (1993)
6. Stanek, K.Z., et al., *ApJ* (in press: astro-ph/0304173)
7. Frail, D.A., et al., *ApJL* **562**, L55 (2001)
8. Salmonson, J. D., & Galama, T. J., *ApJ* **569**, 682 (2002)
9. Zhang, B., & Meszaros, P., *ApJ* **581**, 1236 (2002)
10. Rossi, E., Lazzati, D., & Rees, M.J., *MNRAS* **332**, 945(2002)
11. Woosley, S.E., & MacFadyen, A.I. *A&AS* **138**, 499 (1999)
12. Ioka K., & Nakamura T., *ApJ* **554**, L163 (2001)
13. Lyutikov, M., & Blandford, R., "Electromagnetic Outflows in GRBs," in *Beaming and Jets in Gamma Ray Bursts*, Copenhagen, 2002, in press
14. Fryer, C.L., Holz, D.E., & Hughes, S.A., *ApJ* **565**, 430, (2002)
15. Blondin, J.M., Mezzacappa, A., & DeMarino, C., *ApJ* **584**, 971 (2003)
16. MacFadyen, A. I., Woosley, S. E., & Heger, A., *ApJ* **550**, 410 (2001)
17. Fryer, C.L., & Meszaros, P., ApJ submitted (astro-ph/0303341)
18. Young, T.R., Baron, E., & Branch, D., *ApJ* **449**, L51 (1995)
19. Reeves, J.N., et al., *Nature* **416**, 512 (2002)
20. Wang, L., & Wheeler, J.C., "Supernova Are Not Round", in Sky & Telescope, January 2002, p. 40
21. Wang, L., Howell, D.A., Hoflich, P., & Wheeler, J.C., *ApJ* **550**, 1030 (2001)
22. Coburn, W. & Boggs, S.E., *Nature* **423**, 415 (2003)
23. Lyutikov, M., Pariev, V.I., & Blandford, R., *ApJ* submitted (astro-ph/0305410)
24. Norris, J.P., Marani, G.F., & Bonnell, J.T., *ApJ* **534**, 238 (2000)
25. Norris, J.P., *ApJ* **579**, 386 (2002)
26. Fenimore, E.E., & Ramirez-Ruiz, E., preprint (astro-ph/9906125)
27. Reichart, D.E., et al., *ApJ* **552**, 57 (2001)
28. Lloyd-Ronning, N.M., & Ramirez-Ruiz, E., *ApJ* **576**, 101 (2002)
29. Norris, J.P., Scargle, J.D., & Bonnell, J.T., in *Proc. of Gamma 2001*, AIP **587**, ed. S. Ritz, N. Gehrels, & C.R. Shrader, New York: AIP, 2001, p. 176
30. Lamb, D.Q., & Reichart, D.E., *ApJ* **536**, 1 (2000)
31. Bloom, J. S., et al., *ApJ* **506**, L105 (1998)
32. Salmonson, J.D., *ApJ* **546**, L29 (2001)
33. Band, D.L. private communication
34. Nakamura, T., *ApJ* **522**, L101 (1999)
35. Kulkarni, S. R., et al., *Nature* **395**, 663 (1998)
36. Fritchel, P., "The Second Generation LIGO Interferometers," in *Astrophysical Sources for Ground-based Gravitational Wave Detectors*, AIP **575**, ed. J. M. Centrella, New York: AIP, 2001, p. 15